\documentclass[aip,jcp,floatfix,twocolumn,showpacs,10pt]{revtex4-1} 
\usepackage{color,soul} 
\usepackage{graphicx}
\usepackage{amsmath, amsthm, amssymb}
\usepackage{amsthm}
\usepackage{multirow} 
\usepackage[normalem]{ulem}
\newcommand{\be}{\begin{equation}}
\newcommand{\ee}{\end{equation}}
\newcommand{\bea}{\begin{eqnarray}}
\newcommand{\eea}{\end{eqnarray}}
\begin{document} 
\title{Aggregation Dynamics of Rigid Polyelectrolytes} 
\author{Anvy Moly Tom} 
\email{anvym@imsc.res.in} 
\affiliation{The Institute of Mathematical Sciences, C.I.T. Campus, 
Taramani, Chennai 600113, India} 
\author{R. Rajesh} 
\email{rrajesh@imsc.res.in}  
\affiliation{The Institute of Mathematical Sciences, C.I.T. Campus, 
Taramani, Chennai 600113, India} 
\author{Satyavani Vemparala} 
\email{vani@imsc.res.in} 
\affiliation{The Institute of Mathematical Sciences, C.I.T. Campus, 
Taramani, Chennai 600113, India} 
\date{\today}

\begin{abstract}
Similarly-charged polyelectrolytes are known to attract each other and 
aggregate into bundles when the charge density of the polymers exceeds a 
critical value that depends on the valency of the counterions. The dynamics of aggregation
 of such rigid polyelectrolytes are studied using 
large scale molecular dynamics simulations.  We find that the morphology 
of the aggregates depends on the value of the charge density of the 
polymers. For values close to the critical value, the shape of the
aggregates is cylindrical with height equal to the length of a single 
polyelectrolyte chain. However, for larger values of charge, the linear 
extent of the aggregates increases as more and more polymers 
aggregate.In both the cases, we show that the number of aggregates 
decrease with time as power laws with exponents that are not numerically 
distinguishable from each other, and are independent of charge density 
of the polymers, valency of the counterions, density, and length of the 
polyelectrolyte chain. We model the aggregation dynamics using the 
Smoluchowski coagulation equation with kernels determined from the 
molecular dynamics simulations, and justify the numerically obtained 
value of the exponent. Our results suggest that, once counterions 
condense, effective interactions between polyelectrolyte chains 
short-ranged and the aggregation of polyelectrolytes is 
diffusion-limited.
\end{abstract}

\maketitle

\section{Introduction} 

Many biologically relevant polymers such as DNA, actin and microtubules, 
have charged, rigid or semiflexible backbone structures, 
and may aggregate into bundles in the presence of 
counterions~\cite{Bloomfield, Needleman, Claessens, Huber-actin, 
Sarah-actin,Angelini03}. The aggregates of such 
biological polymers play an important role in cell scaffolding and 
possess superior mechanical 
properties compared to well known synthetic 
flexible polymers~\cite{Fletcher,Stricker}. More recently, it has been 
possible to synthesize non-biological polymers with comparable backbone 
stiffness and  the ability of these 
polymers to aggregate is an important parameter in using them as 
functional biomimetic materials~\cite{Kouwer, Bertrand}. With recent 
studies on various biological phenomena such as DNA packaging, 
cytoskeletal organization, understanding counterion mediated aggregation 
of charged polymers becomes very relevant~\cite{Janmey, Wong-Rev}.

The aggregation of rigid polyelectrolyte (PE) chains has been studied 
extensively via experiments~\cite{Bloomfield,tang,sedlak,tanahatoe, 
borsali,zribi,butler,bordi,Muhlrad,huang2014}, 
simulations in the presence and absence of salt/solvent~\cite{Fazli1,Fazli2,savelyev,sayar1,sayar2,stevens1,stevens2,pietronave,jensen,luan,bruinsma,diehl2001} 
and theoretical approaches~\cite{ray, 
perico2,Zhou-DFT,ermoshkin,Broedersz,perico2,manning3,yethiraj97,yethiraj09,manning2014}. 
While it has been fairly 
well established that multivalent counterions induce aggregation among 
the similarly charged PEs, the ability of monovalent counterions to 
induce a similar aggregation is
debated~\cite{sedlak,tanahatoe,borsali,ermi,zhang,ray,huang2014,perico2,zribi,butler,arenzon1,arenzon2,solis,jensen,stevens1,stevens2,diehl2001,allahyarov,savelyev}.
We recently showed, using 
molecular dynamics simulations as well as computing potential of mean 
force, that for high enough charge density along PE 
backbone, monovalent counterions also induce aggregation~\cite{Anoop}.  Further, this 
critical charge density for aggregation was numerically shown to be 
equal to the critical charge density for the extended-collapsed 
transition of a {\it flexible} polyelectrolyte chain~\cite{Anoop,varghese2011}.

While the aspect of attraction between 
similarly-charged PE chains, typically using coarse-grained
bead-spring models,  is numerically well-studied in the literature ~\cite{Fazli2,savelyev,sayar1,sayar2,stevens1,stevens2,pietronave,jensen,luan,bruinsma}, 
the dynamics of aggregation of such PE chains is 
less studied. This is in part due to the computational cost of simulating large 
number of PE chains with long-ranged Coulomb interactions. Using a 
hybrid Monte Carlo scheme, and simulating a system with 61 PE chains, it 
was argued that for intermediate values of the charge density, 
finite-size PE bundles exist at thermodynamic equilibrium, while further 
increase of charge density, results in phase separation and 
precipitation~\cite{sayar1,sayar2}. Using similar parameters, the 
temporal dependence of the number of clusters of different sizes were 
obtained in Refs.~\cite{Fazli2,Fazli1}. The numerical data was modeled 
by the Smoluchowski coagulation equation which is the basis of classical 
mean-field model of understanding aggregation 
kinetics~\cite{Smoluchowski}, and the number of clusters may be deduced
to decrease with time as $t^{-1}$. However, the coagulation
kernel  was one for particles with
equal masses and sizes. This assumption seems unreasonable as the
aggregate sizes become heterogeneous with time.
In addition, it is not very clear how 
parameters such as the valency of the counterions, the charge density of 
PE chains, or the overall number density of the system, affect the 
aggregation dynamics.

In this paper, using molecular dynamics (MD) simulations (model and MD 
details in Sec.~\ref{sec:model}), we demonstrate that the aggregation of 
similarly charged rigid PEs is independent of linear charge density of 
the polymer chains (higher than a critical value required for onset of 
aggregation) and valency of counterions, and that the number of
aggregates decrease in time as a power law $t^{-\theta}$ where
$\theta=0.62 \pm 0.07$  (Sec.~\ref{sec:theta}).  
The coagulation process is modeled using Smoluchowski
equation with a coagulation kernel determined from the MD simulations
and reproduces the numerically obtained value for $\theta$, implying
that aggregation is  diffusion-limited and  primarily 
driven by short-range interactions (Sec.~\ref{sec:smoluchowski}). 
We find that two merging aggregates approach either perpendicular to
each other or in a collinear manner depending on the charge density
(Sec.~\ref{sec:merge}). Section IV contains a summary and discussion.

\section{Methods \label{sec:model}} 

We consider a system of $N_r$ rigid PE chains. Each PE chain  
consists of $N_m$ monomers, of charge $+qe$, 
connected by bonds. The counterions have 
charge $-Zqe$, where $Z$ is the valency of the 
counterion. In this paper, we consider $Z= 2,3$ corresponding to 
divalent and trivalent counterions respectively. The number of 
counterions are chosen such that the system is overall charge neutral. 
The interactions between the particles are described below:

{\it Excluded volume}: The excluded volume interaction is 
modeled by the 6-12 Lennard Jones potential:
\begin{equation}
U_{LJ}(r_{ij})=4\epsilon _{ij} \left[\left(\frac{\sigma_{ij}}{r_{ij}}\right)^{12}-\left(\frac{\sigma_{ij}}{r_{ij}}\right)^6\right],
\end{equation}
where where $r_{ij}$ is the distance between particles $i$ and $j$,
$\epsilon _{ij}$ is the minimum of the potential and $\sigma_{ij}$ 
is the inter-particle distance at which the potential is zero. 
Both $\epsilon_{ij}$ and $\sigma_{ij}$ are (in reduced 
units) set to $\epsilon_{ij}=\epsilon=1.0$ and $\sigma_{ij}=\sigma=1.0$
for all pairs of particles. The Lennard Jones 
potential is chosen to be zero beyond a cut-off distance
$r_c=\sigma$, such
that the  excluded volume interaction between all pairs is purely repulsive. 

{\it Coulomb}: The electrostatic interaction 
is
\begin{equation}
U_{c}(r_{ij})=\frac{q_iq_j}{4\pi\epsilon_0 r_{ij}},
\label{eq.1}
\end{equation}
where $q_i$ and $q_j$ are the charges of $i^{\text{th}}$ and 
$j^{\text{th}}$particle, and $\epsilon_0$ is the permittivity.

{\it Bond stretching}:  The nearest-neighbor monomers 
along the PE chains are connected by harmonic springs: 
\begin{equation}
U_{bond}(r_{ij})=\frac{1}{2}k(r_{ij}-b)^2,
\end{equation}
where $k$ is the spring constant and $b$ is the equilibrium bond length. 
We set $b=1.12 \sigma$ and $k=500.0$. 
 
{\it Bond bending}: To model rigid PEs, a bond bending 
potential is introduced between two adjacent bonds:  
\begin{equation}
U_\theta(\theta)=k_\theta[1+\cos\theta],
\end{equation}
where $\theta$ is the angle between the bonds.  The strength of this 
interaction is set to a large value $k_\theta =1000.0$. 

The linear charge density along the PE chain is parameterized by a 
dimensionless quantity $A$:
\begin{equation}
A=\frac{q^{2}\ell_{B}}{b},
\label{eq.4}
\end{equation}
where $\ell_{B}$ is the Bjerrum length, the length scale below which 
electrostatic interactions dominate thermal energy~\cite{Russel}
\begin{equation}
\ell_{B}=\frac{e^{2}}{4\pi\epsilon_0 k_{B}T},
\label{eq.5}
\end{equation}
where $k_{B}$ is the Boltzmann constant and $T$ is temperature.

All the simulations  are performed for $N_r=100$ PE chains at values 
of $A$ that are larger than the critical value beyond which the PEs aggregate, as 
determined in Ref.~\cite{Anoop}. A variety of parameters 
such as $A$, valency of the counterions, PE chain length and density of the 
system are varied and the details of the systems simulated are given 
in Table~\ref{table1}. The analyses are performed over $20$ initial conditions
for each set of parameter values in Table~\ref{table1}. 
\begin{table}
\caption{\label{table1} The different values of valency of counterions ($Z$), 
charge density $A$,  PE chain length ($N_m$), and density used in the MD simulations.
Density is
expressed in terms of 
$\rho =3.8 \times 10^{-4}$. The analyses are performed over $20$
initial conditions
for each set of parameter values.
}
\begin{ruledtabular}
\begin{tabular}{llll}
$Z$&$A$& $N_m$& Density \\
\hline
  &$2.01$&30&$\rho$\\
  &$3.57$& $30$&$0.75\rho$\\
  &$3.57$&$30$&$\rho$\\
  &$3.57$&$30$&$1.5\rho$\\ 
$3$  &$3.57$&$30$&$2\rho$\\
  &$4.52$&$30$&$\rho$\\
  &$5.57$&$30$&$\rho$\\
   &$6.75$&$30$&$\rho$\\
  &$8.03$&$30$&$\rho$\\
 \hline
 &$3.57$&$15$&$\rho$\\
  &$3.57$&$30$&$\rho$\\
 $2$ &$3.57$&$60$&$\rho$\\
  &$5.58$&$30$&$\rho$\\
  &$14.28$&$30$&$\rho$\\
\end{tabular}
\end{ruledtabular}
\end{table}

The equations of motion are integrated in time using the 
molecular dynamics simulation package LAMMPS~\cite{lammps1,lammps2}. The 
simulations are carried out at constant temperature (T=1.0), maintained 
through a Nos\'{e}-Hoover thermostat (coupling constant 
$=0.1$)~\cite{nose,hoover}. The long-ranged 
Coulomb interactions are evaluated using the particle-particle/particle-mesh (PPPM) 
technique~\cite{hockney}. The time step for integrating equations of motion is chosen as 
$0.001$. A homogeneous initial state is prepared as follows.
$N_r$ non-overlapping PE chains of length $N_m$ are placed in a cubic box with periodic 
boundary conditions with randomly distributed counterions. The charge density of
the PE chains is set to a very small value ($A=0.22$) which ensures that counterions do not
condense onto the PE chains.  The system is then evolved to ensure homogeneous
distribution of the PE chains and the counterions.
Twenty random  configurations, which are temporarily well separated, are chosen and appropriate 
values of $A$ are chosen for further simulations.

\section{Results}

\subsection{Aggregation Dynamics \label{sec:theta}}

We first present results for the temporal dependence of number of
aggregates.  Two PEs are said to form an aggregate of size two if the
distance between any two monomers (not from the same PE) is
less than $2 \sigma$, and the same definition is extended to an
aggregate of size $m$.
\begin{figure}
\includegraphics[width=\columnwidth]{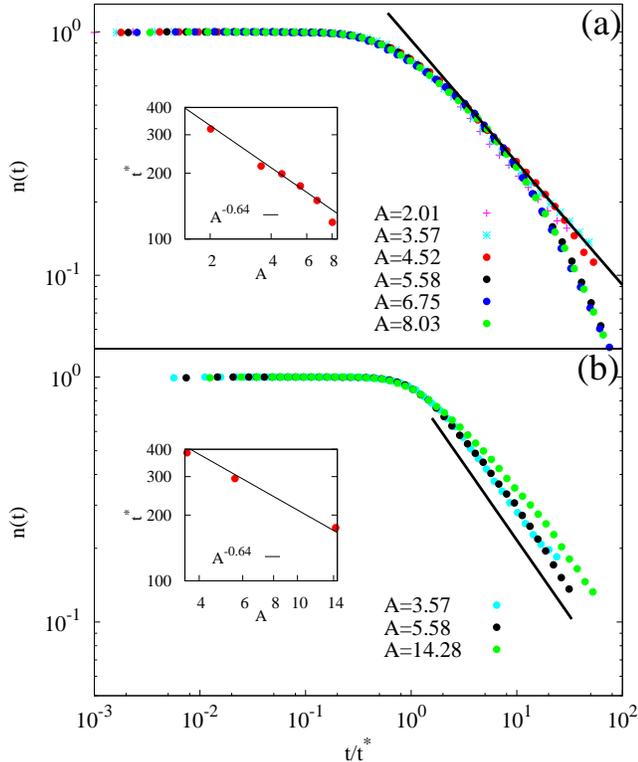}
\caption{The variation of the fraction of aggregates $n(t)$ with
scaled time $t/t^*$ for different values of A for systems with (a)
trivalent and (b) divalent counterions.
$t^*$ is the time at which $n =0.9$. The straight lines are power law $(t/t^*)^{-0.62}$.
Insets shows the dependence of $t^*$ on $A$. }
\label{fig:nversust}
\end{figure}

Figure~\ref{fig:nversust}(a) and (b) shows the fraction of aggregates
$n(t)=N(t)/N(0)$, as a function of time $t/t^*$ for trivalent and
divalent counterions respectively. Here, $N(t)$ is the number of
aggregates at time $t$, and $t^*$ is  the time taken for number of
clusters to reach $90\%$
of $N(0)$. For small times $n(t) \approx 1$ and aggregation is initiated 
only after $t^*$, which is the
time taken for counterion condensation to occur. Beyond $t^*$,
$n(t)$ decreases as a power law $t^{-\theta}$. 
From Fig.~\ref{fig:nversust}, we find that the exponent $\theta$
is independent or utmost very weakly dependent on $A$ as well as
valency, and has the value 
$\theta = 0.62\pm 0.07$. 

\begin{figure}
\includegraphics[width=0.7\columnwidth]{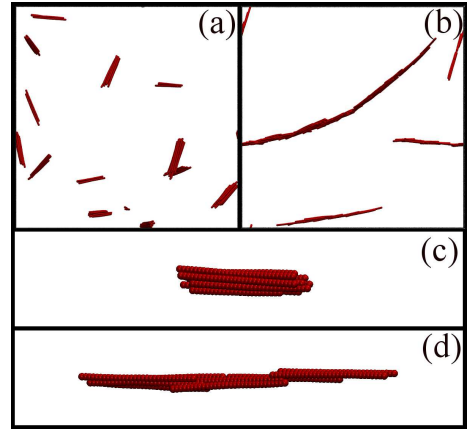}
\caption{Snapshots of systems with trivalent counterions for
(a)$ A=2.01$ and (b) $A=8.03$. In (c) and (d), magnified images of
clusters of size 10 in the snapshots of (a) and (b) are shown.
(Counterions are not shown  in the picture for the sake of clarity.)} 
\label{fig:snapshot_cluster}
\end{figure}

For systems with trivalent
counterions and large $A$, we find that $n(t)$ deviates from the power law 
behavior at long times [see Fig.~\ref{fig:nversust}(a)].
To understand this crossover, 
we study the morphology of the aggregates.
Figure~\ref{fig:snapshot_cluster} shows snapshots of the system for
$A=2.01$ and $A=8.03$ in (a) and (b), along with enlarged snapshots of 
aggregates of size
$10$ in (c) and (d). For smaller values of $A$, the aggregates are
cylindrical in shape with length of the aggregate being roughly the same as the
length of a PE, while for larger $A$, the aggregates are linear but with 
larger aggregates having longer length.
The crossover seen in Fig.~\ref{fig:nversust}(a) at long times occurs
only for aggregates  whose length increases with aggregate size, 
and is likely a finite size effect 
due to the size of the aggregate becoming comparable to the system
size.

\begin{figure}
\includegraphics[width=\columnwidth]{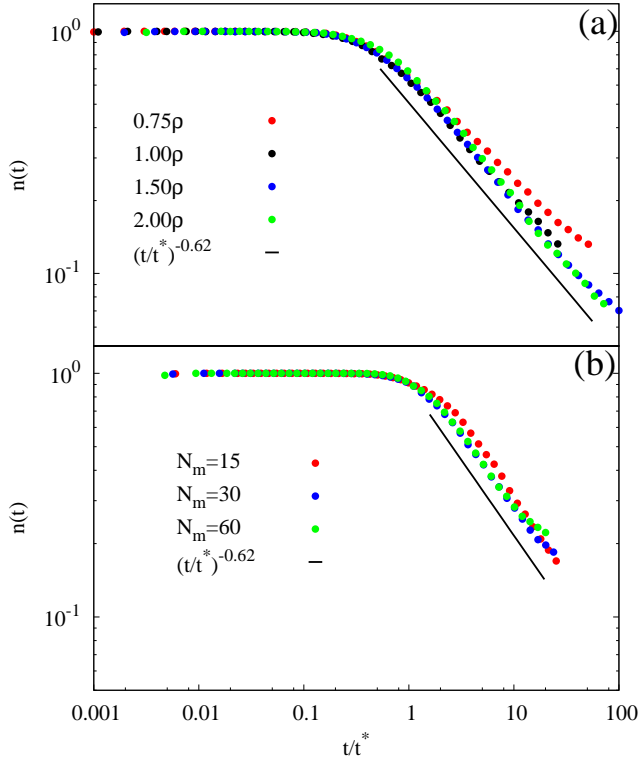}
\caption{Fraction of clusters, $n(t)$, as a function of time 
for (a) different densities and trivalent counter ions, and (b) different PE length  
$N_m$ and divalent counter ions. The data are for $A=3.57$.}
\label{fig:nversust_rho_Nm}
\end{figure}

We also confirm that the exponent $\theta$ 
does not depend on the density $\rho$ as well as $N_m$, the length of the PE chain,
as can be seen
from the collapse of the  data for
different $\rho$ and $N_m$ onto one curve [see
Fig.~\ref{fig:nversust_rho_Nm}(a) and (b)].
We thus conclude that the exponent $\theta$ characterizing the power
law decay of number of aggregates is quite universal and does not
appear to depend on parameters such as valency, density or length of the
PE. It is thus plausible that aggregation is driven by diffusion and
irreversible aggregation (we do not see any fragmentation event) due to 
short-ranged attractive forces. With this assumption, we recast the
aggregation dynamics of PE in terms of the
Smoluchowski equation in Sec.~\ref{sec:smoluchowski}.

\subsection{Recasting results in terms of Smoluchowski equation \label{sec:smoluchowski}}

The Smoluchowski equation describes irreversible aggregation of
particles that are transported by some mechanism such as
diffusion or ballistic motion. 
In Sec.~\ref{sec:theta}, we showed that the PE aggregation dynamics
is independent of PE charge density ($A$) and valency of counterions ($Z$). 
We model the aggregation as one of diffusing neutral rod-like
particles that aggregate due to short-ranged attraction.

The Smoluchowski equation for irreversible aggregation (for reviews, see~\cite{leyvraz2005,ccareview}
is 
\begin{align}
\frac{dN(m,t)}{dt}&= \frac{1}{2}\sum_{m_1=1}^{m-1} K(m_1,m-m_1) N(m_1)
N(m-m_1) \nonumber \\
& -\sum_{m_1=1}^\infty K(m,m_1) N(m) N(m_1),
\label{eqn1}
\end{align}
where $N(m,t)$ is the number of aggregates of size $m$ at time $t$,
and $K(m_1,m_2)$ is the rate at which two masses $m_1$ and $m_2$
collide. The first term in Eq.~\eqref{eqn1} describes the 
aggregation of particles to form an aggregate of size $m$, while the
second term  describes the loss of an aggregate of size $m$ due to
collision with another aggregate.

If the kernel $K(m_1,m_2)$ is a homogeneous function of its arguments
with homogeneity exponent $\lambda$, i.e., $K(hm_1,h m_2) = h^\lambda
K(m_1,m_2)$, then the number of aggregates $N(t) = \sum_m N(m,t)$,
decreases in time as a power law $N(t) \sim t^{-\theta}$, where
\be
\theta=\frac{1}{1-\lambda}, \quad \lambda<1.
\label{eq:beta}
\ee

To construct the kernel $K(m_1,m_2)$, we consider the aggregates to be
effective spheres of radius $\sqrt{\ell^2+r^2}$, where $\ell$ and $r$
are the the height and radius of the cylindrical aggregate. This is
justified because we observe that the aggregates rotate at a rate that is 
much larger
than the rate of collision (see Supplementary
material)
For diffusing spheres in three
dimensions, the coagulation kernel is known to be (for example, see~\cite{krapivsky2010kinetic})
\begin{equation}
K(m_1,m_2) \propto [D(m_1)+D(m_2)][R(m_1)+R(m_2)],
\label{eq}
\end{equation}
where $D(m)$ and $R(m)$ are the diffusion constant and effective radius of an
aggregate of $m$ PEs.
In the absence of a solvent, the diffusion constant is inversely
proportional to its mass:
\be
D(m) \propto m^{-1}.
\ee

\begin{figure}
\includegraphics[width=\columnwidth]{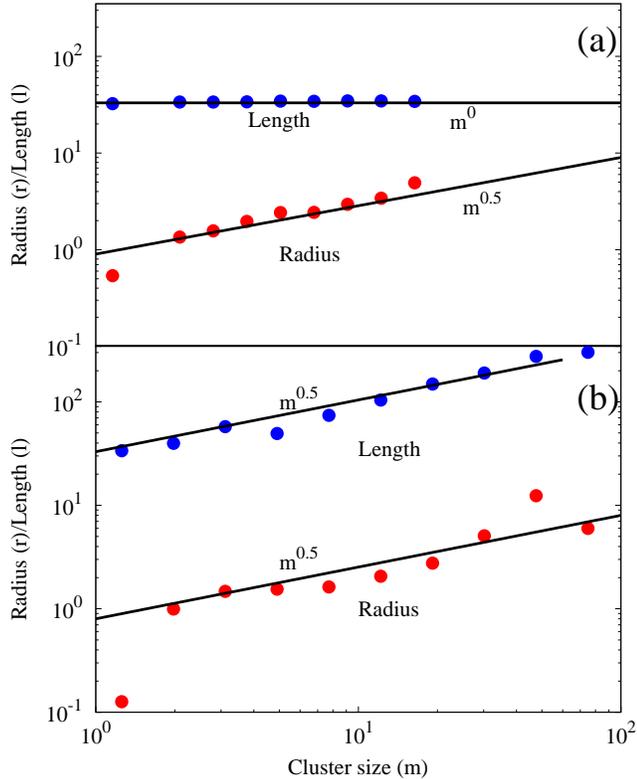}
\caption{The length and radius of different cluster sizes $m$ for (a) $A=2.01$ and (b) $A=8.03$.
 }
\label{fig:LnR}
\end{figure}

The dependence of the radius $R(m)$ on $m$ may be determined by  
studying the geometry of the aggregates obtained from the MD
simulations. The geometry of  an aggregate may be quantified by the 
the eigenvalues of the gyration tensor $S$ whose elements are
\begin{equation}
S_{\alpha \beta}=\frac{1}{N} \sum _{i=1} ^N r_{i \alpha}r_{i\beta},~\alpha, \beta=1,2,3,
\end{equation}
where $r_{i \alpha}$ is the $\alpha$\textsuperscript{th} component of position vector
$\overrightarrow{r _i}$ of i\textsuperscript{th} particle measured from the center of mass. 
Let the eigenvalues be denoted by
$\lambda_1$, $\lambda_2$, and $\lambda_3$, where 
$\lambda_1 \geq \lambda_2\geq \lambda_3$. Modeling the shape of the aggregate
as a cylinder, we obtain the length and radius of the
aggregate to be  $\ell =  \sqrt{12 \lambda_1}$ and 
$r=\sqrt{2(\lambda_2+\lambda_3)}$. The length and radius, thus measured, are shown in
Fig.~\ref{fig:LnR} for $A=2.01$ and $A=8.03$. For small values of $A$,
$\ell$ is independent of aggregate size $m$, i.e. $\ell \sim m^0$,
while the radius $r$ increases with $m$ as $r \sim \sqrt{m}$ [see
Fig.~\ref{fig:LnR}(a)]. For large values of $A$, we find that $\ell
\sim \sqrt{m}$ and $r\sim \sqrt{m}$ [see 
Fig.~\ref{fig:LnR}(b)]. Thus,  aggregation is controlled by two types of kernels:
\be
\frac{K(m_1,m_2)}{m_1^{-1}+m_2^{-1}} \propto \begin{cases} \sqrt{N_m^2 +m_1}+ \sqrt{N_m^2 +m_2}
&\mbox{if } A \gtrsim A_c  \\ 
\sqrt{m_1}+ \sqrt{m_2}  & \mbox{if }  A \gg A_c, \end{cases} 
\ee
where $A_c$ is the critical charge density beyond which aggregation sets in.
The radius and length of cluster are calculated from the eigenvalues of 
gyration tensor.

For $A \gg A_c$, the kernel is homogeneous with homogeneity exponent $\lambda=-1/2$. From
Eq.~\eqref{eq:beta}, we obtain $\theta=2/3$. This is in excellent agreement with the numerical
value of $0.62\pm0.07$ from molecular simulations. When $A \gtrsim A_c$, the kernel is no longer
homogeneous. For large $m_1$ and $m_2$, it is homogeneous with $\lambda=-1/2$. 
Equation~\eqref{eq:beta} gives $\theta=2/3$. On the other hand, for small $m_1$, $m_2$,
we may ignore the dependence of radius on mass, and the kernel is homogeneous with 
$\lambda=-1$ or equivalently $\theta=1/2$. The numerically obtained value of $0.62\pm0.07$ lies
between these two bounds $0.5$ and $0.67$.

\begin{figure}
\includegraphics[width=\columnwidth]{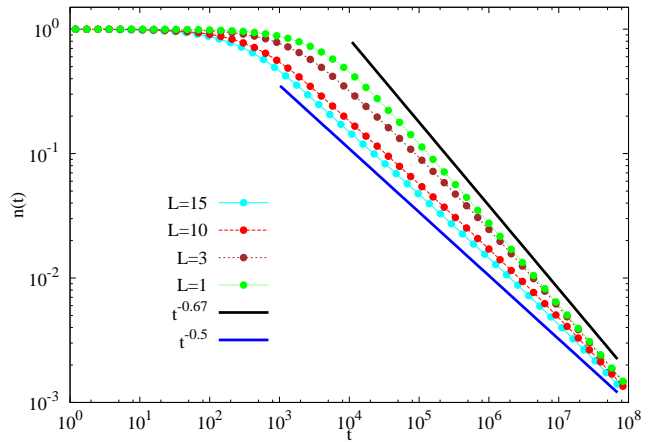}
\caption{Monte Carlo simulations with kernel as in Eq.~\eqref{eq:mckernel} for different
values of the parameter $L$.}
\label{fig:mc}
\end{figure}

In our MD simulations, computational expense limits the number of PEs that can be studied to few hundreds. However,  large-scale Monte Carlo simulations
can be used to study the effect of the kernel for $A \gtrsim A_c$ on
the measured $\theta$. In these simulations, we start with $M=10^5$
particles of mass $1$. Any pair of particles of masses $m_1$ and $m_2$ undergo aggregation
to form a particle of mass $m_1+m_2$ with rate
\be
K(m_1,m_2)=\Lambda ({m_1^{-1}+m_2^{-1}}) (\sqrt{L^2 +m_1}+ \sqrt{L^2 +m_2}),
\label{eq:mckernel}
\ee
where $L$ is a parameter and $\Lambda$ is chosen to be proportional to $M^{-2}$. The stochastic
processes were simulated using standard Monte Carlo techniques. Each parameter value was
averaged over $1000$ histories.  The results for $n(t)$ for different parameter values are shown in
Fig.~\ref{fig:mc}. As $L$ increases the effective power law changes from $-0.67$ to
$-0.5$, and $\theta=0.62 \pm 0.07$ from molecular dynamics falls within this range.

From the above analysis based on Smoluchowski equation, we obtain $\theta=2/3$ very similar to the value obtained through our MD simulations ($\theta=0.62 \pm 0.07$). 
More accurate determination of $\theta$ through MD simulations will require much larger systems to be simulated for much longer times, currently a very expensive proposition.
In earlier simulations of rigid PEs \cite{Fazli2}, it was suggested that the decay of the number of aggregates scales with time as $t^{-1}$ different from the exponent obtained in this work 
($t^{-0.62}$). This difference could be attributed to the assumption made in the analysis based on Smoluchowski equation in the earlier paper \cite{Fazli2} that collisions occur between aggregates
of approximately equal size \cite{Russel}. In this work, we explicitly take into consideration collisions between aggregates of different sizes, which is much more realistic picture and hence we consider the result
obtained in this work to be more accurate.

\subsection{Two kinds of dynamics \label{sec:merge}}

\begin{figure}
\includegraphics[width=\columnwidth]{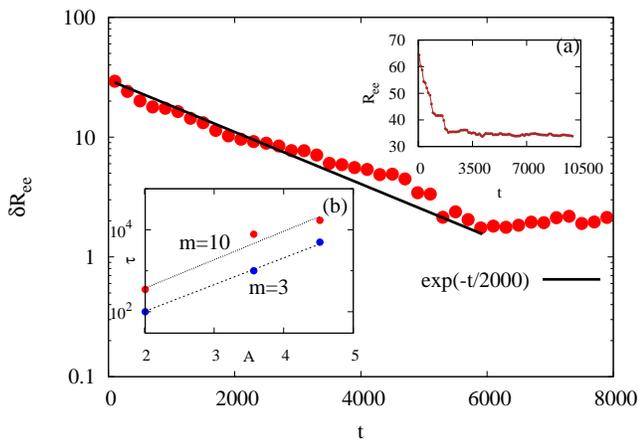}
\caption{The deviation of the end to end distance of the aggregate, 
$\delta R_{ee}$, from its equilibrium value as a function of time $t$. 
It decreases with time as an exponential. The data is for an
aggregate of size three with trivalent counterions, $A=3.57$ and
averaged over three realizations. 
(a) The end to end distance $R_{ee}$ for a single realization for the 
same parameters as in main plot. (b) The variation
of the relaxation times $\tau$ with $A$ for different aggregate sizes.
The straight lines are $\exp(1.51 A)$ ($m=3$) and $\exp(1.61 A)$
($m=10$).}
\label{fig:e2e_dist}
\end{figure}

For large values of $A$, we observed that the aggregates are collinear with the effective length increasing
with size of aggregate[see Fig.~\ref{fig:LnR}(b)].  However, we find that such aggregates, 
when isolated, rearrange themselves from elongated to more compact cylindrical structures whose
lengths are comparable to that of a single PE chain. To quantify this, we extract aggregates of size
$3$ and $10$ from the simulations for $A=8.03$ and with trivalent counterions, isolate them,
and allow them to evolve for different values of $A$. A typical time profile of the end  to end distance,
$R_{ee}$ 
is shown in Fig.~\ref{fig:e2e_dist}(a). It decreases in steps with sudden decreases in length
due to re-arrangement, separated in time. From the history averaged data (see Fig.~\ref{fig:e2e_dist}),
a relaxation time $\tau$ associated with the rearrangement may be extracted.

Thus, there are two time scales in the problem:
one is the diffusion time scale corresponding to the time taken for two aggregates to be
transported nearby, and the second is the sliding time scale $\tau$ corresponding
to the time taken for an aggregate to re-align itself into a compact
cylindrical shape. The sliding time scale increases rapidly with $A$ as seen in 
Fig.~\ref{fig:e2e_dist}(b). For large $A$, the sliding time scale is much larger than the diffusion time scale and the re-alignment
may be ignored.

\begin{figure}
\includegraphics[width=\columnwidth]{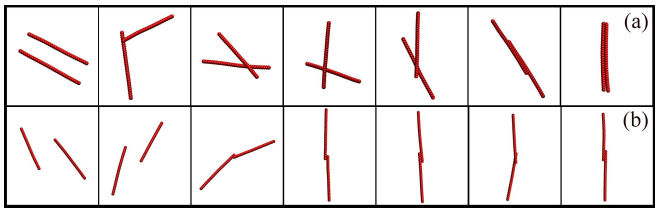}
 \caption{Snapshots describing the merging of two PEs for (a)$A=2.01$ (b)$A=8.03$ for
a system with  trivalent counterions}
\label{fig:formation_cluster}
\end{figure}

We also find that the process by which two aggregates  merge are different for small and large $A$. 
For small $A$, when two polyelectrolytes merge, they first orient in orthogonal directions, and the
point of intersection moves towards the center.  At later times, they align and rearrange themselves from elongated to more compact cylindrical structures [see Fig~\ref{fig:formation_cluster}(a)]. 
For large $A$, the aggregates intersect and align themselves without
sliding [see Fig.~\ref{fig:formation_cluster}(b)].

\section{Discussion and Conclusion}

In this paper, we studied 
the dynamics of aggregation of similarly-charged 
rigid PE chains using extensive MD simulations.
It was shown that the dynamics of 
aggregation is effectively determined by short-ranged interactions
between the PE chains, 
even though the monomers and counterions interact via long-ranged Coulomb 
interactions. We also showed that the number of aggregates decreases with
time as a power law, $t^{-\theta}$, where the exponent $\theta$ is independent of 
the charge density of the PE chains, whether the counterions are divalent or
trivalent, 
number density, and length of the PE chains. The data is modeled using
Smoluchowski equation with coagulation kernel determined from the
MD simulations. From the molecular dynamics simulations, we estimate
$\theta=0.62 \pm 0.07$, which is consistent with the value
$\theta=2/3$ obtained from the Smoluchowski equation.

The current simulations are only for systems with divalent and trivalent counterions. In
an earlier paper~\cite{Anoop}, we had shown 
that monovalent 
counterions induce aggregation among similarly-charged PE chains,
and preliminary data suggested
$\theta \approx 0.66$. This, being consistent with
the results obtained in this paper for divalent and trivalent counterions,
we conclude that the dynamics is independent of valency of
counterions.
The charge density required for aggregation with monovalent
counterions is much larger than that for divalent and trivalent
counterions, resulting in much longer simulations needed for obtaining
good data.  For efficient computational purposes, we restrict the 
simulations in this paper to divalent and trivalent counterions. 

In earlier simulations of rigid PE chains~\cite{Fazli2}, by modeling the data with
the Smoluchowski coagulation equation, it can be deduced that $\theta=1$,
different from $\theta\approx 2/3$ obtained in this paper.  
This difference could be attributed to the assumption made in the analysis 
of Ref.~\cite{Fazli2} that all
aggregates are approximately of same size. In this paper, we explicitly 
take into consideration collisions between aggregates of different sizes, 
which is a much more realistic picture given the heterogeneous 
aggregate size distribution. In addition, the extensive
MD simulations performed in this paper allow us to clearly
distinguish between the exponents $1$ and $2/3$,
and hence we consider the result
obtained in this paper to be more accurate. 

\begin{figure}
\includegraphics[width=\columnwidth]{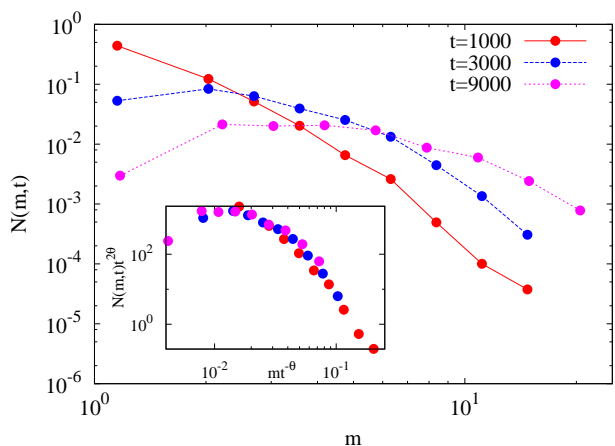}
 \caption{Aggregate size distribution $N(m,t)$ for different times $t$ 
 for a system with trivalent counterions and $A=2.01$. Inset: The data for
 different times collapse onto a single curve when $m$ and $N(m,t)$
 are scaled as shown in the figure with $\theta=0.62$.}
\label{fig:no_bundle}
\end{figure}

It has been argued that 
for intermediate values of the charge density, 
finite size PE bundles exist at thermodynamic equilibrium, while further 
increase of charge density, results in phase separation and 
precipitation~\cite{sayar1,sayar2,Fazli1,Fazli2}. However, in our
simulations, for all the values of charge densities that we have
considered, the number of aggregates decrease continuously as a power
law, and shows no tendency to plateau which would be the case if
finite sized bundles at thermodynamic equilibrium existed. In addition, we find that
the cluster size distribution for different times obeys a simple scaling 
$N(m,t) \simeq t^{-2 \theta}
f(m t^{-\theta})$, where $f$ is a scaling function (see Fig.~\ref{fig:no_bundle}), 
showing that the system continuously coarsens
to presumably a phase separated state. This discrepancy in results could be 
due to the fact that the observation of finite sized bundles in
Refs.~\cite{sayar1,sayar2,Fazli1,Fazli2} was based on an arbitrarily chosen
equilibration time. 

It has also been suggested in many previous papers that similarly-charged rigid PE chains tend to approach
each other at right angles, align and then slide to align with the bundle while merging with each other. This mode
has been referred in the literature as zipper model~\cite{Nguyen02,sayar2,Fazli2, Fazli1}. Another model of approach called collinear model was also 
proposed for rigid PE chains, in which the centers of mass of approaching PE chains lie on a line parallel to their longer axes~\cite{Nguyen02}. This model
was shown to have lower kinetic barrier of approach and can explain the observation of elongated structures in experiments~\cite{Bordi06, susoff08}.
From our MD simulations, we see that the approach of merging depends significantly on the charge density of rigid PE chains. While zipper model seems to be the 
mode of aggregation for PE chains with lower charge density, the approach mechanism changes to collinear model for PE chains with high charge density. 

 All the simulations performed in this paper were in the
absence of a solvent. From the obtained results, we expect that adding
a solvent will result in modifying the dependence of diffusion
constant on the aggregate size, and the results from the Smoluchowski
equation should be carried forward. Likewise, adding salt will make
the bare interactions even further short-ranged due to screening. This should not
change the results except for modifying the critical charge density
required for the onset of aggregation.

\begin{acknowledgments}
We thank Upayan Baul for helpful discussions.
The simulations were carried out on the 
supercomputing
machines Annapurna, Nandadevi and Satpura at The Institute of Mathematical Sciences.
\end{acknowledgments}

%\bibliography{aggregate} 

%merlin.mbs apsrev4-1.bst 2010-07-25 4.21a (PWD, AO, DPC) hacked
%Control: key (0)
%Control: author (8) initials jnrlst
%Control: editor formatted (1) identically to author
%Control: production of article title (-1) disabled
%Control: page (0) single
%Control: year (1) truncated
%Control: production of eprint (0) enabled
%

\end{document}